\documentclass{article}
\usepackage{arxiv}

\usepackage[utf8]{inputenc} 
\usepackage[T1]{fontenc}    
\usepackage{hyperref}       
\usepackage{url}            
\usepackage{booktabs}       
\usepackage{amsfonts}       
\usepackage{nicefrac}       
\usepackage{graphicx}
\usepackage{doi}

\usepackage{setspace} 
\usepackage{amsmath} 
\usepackage[authoryear, round]{natbib} 
\usepackage{url} 
\usepackage[table]{xcolor} 

\usepackage{geometry}    
\usepackage{float} 
\usepackage{threeparttable}
\usepackage{graphicx}

\title{Public Education Spending and Income Inequality}
\author{ \href{https://orcid.org/0000-0001-8689-6204}{\includegraphics[scale=0.06]{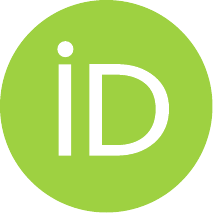}\hspace{1mm}Ishmael Amartey}\thanks{corresponding email: iamartey1@niu.edu} \\
	Department of Economics\\
	Northern Illinois University\\
	Dekalb, IL 60115 \\
}

\date{December 11, 2025}

\renewcommand{\headeright}{\textit{M.A. Research Paper}}
\renewcommand{\undertitle}{}
\renewcommand{\shorttitle}{\textit{Public Economics}}

\hypersetup{
pdftitle={A template for the arxiv style},
pdfsubject={q-bio.NC, q-bio.QM},
pdfauthor={David S.~Hippocampus, Elias D.~Striatum},
pdfkeywords={First keyword, Second keyword, More},
}

\begin{document}
\maketitle
\begin{abstract}
	This paper investigates the relationship between public education spending and income inequality across U.S. counties from 2010–2022 using quantile regression methods. The analysis shows that total per-pupil education spending is consistently associated with a small increase in income inequality, with stronger effects in high-inequality counties. In contrast, the composition of education spending plays a substantially more important role. Reallocating budgets toward instructional, support service, and other current expenditures significantly reduces income inequality, particularly at the upper quantiles of the Gini distribution. Capital outlays and interest payments exhibit weaker and mixed effects. Economic and demographic factors, especially poverty, median income, and educational attainment, remain dominant drivers of inequality. Overall, the results demonstrate that how education funds are allocated matters more than how much is spent, underscoring the importance of budget composition in using public education policy to promote equity.
\end{abstract}


\section{Introduction}

The United States is one of the countries that prioritizes education and spends a great deal of money on it at all levels. As established in various congressional acts such as the Elementary and Secondary Education Act (ESEA), Adult Education and Family Literacy Act (AEFLA), and the Individuals with Disabilities Education Act (IDEA), the federal government is mandated to support educational needs of states by providing funds, so no person of schoolgoing age, no matter their physical or mental challenges, is left out. In many instances the main focus for these grants is on low-income families, low achieving students, and rural schools, as a result, public education has become a pivot for social mobility and a primary tool for mitigating income inequality. The US educational system has a three-linked funding system: federal, state, and local, prompting a critical and urgent question on the impact of these public investments on economic outcomes and to what extent do they actually reduce income inequality.

Over the years, many economists have developed many models that seek to estimate the relationship that between public spending on education and how it really elevates people out of the lower echelon of society. Yet a crucial gap persists: while the level of funding is often debated, much less is known about how the composition of education spending, how dollars are allocated across instructional, operational, capital, and debt-service and how these categories shapes economic inequality. This distinction is essential, as funding totals may mask redistributive inefficiencies or compositional choices that either exacerbate or alleviate income disparities. Without understanding these nuances, policymakers risk advocating for blunt increases in education budgets that fail to achieve equity objectives.

This study seeks to explore the relationship between fiscal policy, educational investment, and the distribution of income using county-level data, advancing understanding of the justification and impact of public education spending on income inequality. It contributes to the literature not only by confirming whether education spending matters for income distribution, but also by highlighting which types of spending are most effective in promoting equity. By employing quantile regression techniques, the study moves beyond average effects to examine how these relationships vary across the full spectrum of inequality—from the most equal to the most unequal counties. In doing so, it provides actionable evidence on for whom and under what conditions education spending reduces inequality, offering policymakers practical guidance on designing budgets that maximize social mobility, reduce disparities, improve economic opportunities, and strengthen the role of public education as a tool for equity.

\section{Literature Review}
\cite{schultz1963}, argues that investments in educational expenditures are investments in human capital that increases individual productivity and earnings, which serves as a powerful tool for creating a more equitable distribution of income. From this perspective, equitable access to and the quality of education should directly reduce income inequality by equipping a larger segment of the population with valuable skills, leading to a compression of the wage distribution.
This was evident in \cite{glomm1992}, when their comparison of public and private education regimes demonstrated that a public educational system leads to a decline in income inequality and higher growth in the long run compared to a private system despite the private system yielding higher per-capita income. Therefore, access to a well publicly funded resource may significantly impact income inequalities.

Similar studies have provided support for this claim. One example is the work of \cite{sylwester2002}, who found that higher public education expenditures (as a share of GDP) are associated with a decline in income inequality, and the effect is larger in high-income nations. This suggested that once a basic level of development is achieved, marginal investments in education are more effectively channeled toward equalization. 

In the case of the U.S., \cite{goodspeed2000} provided further evidence that education spending has an indirect effect on growth by reducing income inequality.
Although many studies have shown the impact of government spending on income inequality, the issue of endogeneity usually arise. Specifically areas with higher inequality or poorer economic conditions will get more educational funding through federal programs as compared to more economic vibrant areas creating a negative correlation. 

This issue is addressed in \cite{jackson2015}. Their study uses school finance reforms (SFRs) from the 1970s as exogenous "shocks" to school spending. They linked the SFRs data to longitudinal data from the Panel Study of Income Dynamics (PSID), to compare the adult outcomes of children who were differentially exposed to funding increases based on their birth year and school district. They found that a 10\% increase in per-pupil spending throughout all 12 school-age years led to significant improvements for low-income children, including 0.43 more years of education, 9.5\% higher wages, and a 6.8 percentage point reduction in the annual incidence of adult poverty. For children from families in high income ranges the effects were smaller and often statistically insignificant. This demonstrates that effects of public education expenditures are heterogeneous, primarily benefiting economically disadvantaged students, which is the exact goal through which education spending would be expected to reduce overall income inequality.  

Another study [\cite{seefeldt2018}] using state-level panel data  found that both primary/ secondary and tertiary education spending has a significant inequality-reducing effect, which she attributes to a "compression effect" that equalizes opportunities.

Even though many studies show strong evidence that more school funding can reduce inequality, there is also another group of research that suggests the story is not that simple. These studies argue that money does not automatically create fairness as the effects of education spending go through many different social and economic pathways, and sometimes the results can even be the opposite of what we expect. 

A good example of this comes from \cite{glomm2003}, who studied how education affects income over several generations. In their model, everyone attends the same public schools, receives the same quality of education, and pays the same tax rate. In other words, everything looks perfectly fair. But their results show something surprising: income inequality can actually get worse before it gets better because what children achieve in school is influenced not only by school quality but also by the home environment, they grow up in. Children with highly educated parents often get more help with schoolwork, spend more time studying, and learn skills at home that school alone cannot provide. Meanwhile, children from less advantaged families may not receive the same level of support, even if they go to the exact same school. So, even with equal school funding, the children of richer families may continue to pull ahead and the gap between rich and poor can widen for a period of time simply because parents pass on advantages that money alone cannot erase. 

The story becomes even more complex when we consider how education spending interacts with the economy over time. \cite{sylwester2000} studies this issue and finds something important: countries with higher income inequality today tend to spend more on public education later on. This makes sense when there’s a big gap between the rich and poor, voters are more likely to support policies that help redistribute opportunities, such as better public schooling. However, \cite{sylwester2000} also found that this increased spending can slow down economic growth in the short term. The likely reason is that raising taxes to fund education can temporarily reduce how much money people have to spend or invest. But in the long run, the picture flips. Once those education investments start paying off as students become more skilled workers, they can boost economic growth and reduce inequality in future years. Because of this, if we look at only one point in time, the relationship between education spending and inequality may appear confusing or even negative. To truly understand what’s happening, we have to zoom out and recognize that education is an investment, and it takes time for the benefits to show up.

An interesting twist in studying spending on education came from \cite{artige2023} when they argue that educational spending can actually increase income inequality. They show that spending more on education doesn't automatically make everyone more equal but can sometimes make the rich even richer depending on what jobs people choose. They argue that for workers at the bottom of the income distribution, gaining more skills can raise wages and improve living standards. This reduces the income gap between workers and managers. However, managers who already benefit from high skill levels may see their wages increase even faster when educational improvements make high-skilled occupations more productive and valuable. In that case, inequality can widen. According to \cite{artige2023}, whether education spending increases or decreases inequality depends on the balance of these competing forces, which varies across different economies.

From a policy perspective, \cite{darvas2020} identifies spending on quality education as one of the fiscal policies that can be a "win-win," simultaneously boosting economic growth and reducing income inequality. He places emphasis on the progressive tax system that funds education especially in the U.S where local property taxes finance a significant share of K–12 education, wealthy communities often invest much more in their schools than poorer communities, reinforcing geographic inequality. However, this "win-win" potential may be systematically undermined by the very inequality it aims to address. 

The findings of \cite{demello2006} tell a different story: more unequal societies tend to spend less, not more, on redistributive government programs. Their research supports the "imperfect markets" hypothesis, which states that high inequality and capital market imperfections can lead to a political breakdown in the consensus for redistributive policies. Thus, high initial inequality leads to lower public investment in equalizing mechanisms like education, which in turn perpetuates and may even exacerbate the original inequality; a pattern \cite{darvas2020} observes in the U.S. school funding system. Therefore, while education spending is a powerful tool for reducing inequality, its implementation is often hampered by the pre-existing structural inequalities within a society.

The various research into the relationship on public education spending and income inequality do not always point to a simple conclusion. We have seen evidence that confirms that investing in education can make a real difference as it can significantly improve the life outcomes of children from low-income backgrounds (Jackson et al., 2015). Also, Broader state-level analyses has consistently link higher education spending with lower levels of inequality (Seefeldt, 2018; Sylwester, 2002). In theory, education is a powerful tool for creating a more equitable society (Darvas, 2020). However, the path is not straightforward. 

Many studies, particularly those showing a correlation at the state level, operate at too broad a scale. A state-level finding can highlight the differences between states, but it fails to capture how inequality in a wealthy suburb differs from that in a rural county or a struggling urban center, and how these local conditions influence school funding. We know that more spending can work, but we have a less clear picture of the specific local contexts where it succeeds or fails and how local inequality itself can impact the funding needed for success.

\section{Methodology}
This study employs quantile regression analysis to examine the heterogeneous relationship between public education spending and income inequality across U.S. counties. Building on the panel data framework while addressing the limitations of mean-based estimation approaches, I estimate four econometric models that examine how educational expenditures at different points in the conditional distribution of income inequality relate to the Gini coefficient, following similar approaches in the literature \citep{sylwester2002, artige2023}.
\subsection{Econometric Models}
The study estimates four quantile regression models at five key points in the conditional distribution of the Gini index:
\begin{align}
Q_{\tau}(Gini_{it}) &= \beta_0(\tau) + \beta_1(\tau)PP\_TOTALEXP_{it} + \boldsymbol{\beta}(\tau)\mathbf{X}_{it} + \epsilon_{it}(\tau) \label{eq:m1} \\
Q_{\tau}(Gini_{it}) &= \gamma_0(\tau) + \sum_{k=1}^{5}\gamma_k(\tau)PP\_COMP_{k,it} + \boldsymbol{\gamma}(\tau)\mathbf{X}_{it} + \epsilon_{it}(\tau) \label{eq:m2} \\
Q_{\tau}(Gini_{it}) &= \delta_0(\tau) + \sum_{m=1}^{3}\delta_m(\tau)PS\_SHARE_{m,it} + \boldsymbol{\delta}(\tau)\mathbf{X}_{it} + \epsilon_{it}(\tau) \label{eq:m3} \\
Q_{\tau}(Gini_{it}) &= \theta_0(\tau) + \sum_{n=1}^{5}\theta_n(\tau)PS\_DETAIL_{n,it} + \boldsymbol{\theta}(\tau)\mathbf{X}_{it} + \epsilon_{it}(\tau) \label{eq:m4}
\end{align}

In these equations, $Q_{\tau}(Gini_{it})$ represents the $\tau$-th conditional quantile of the Gini index for county $i$ in year $t$, where $\tau = \{0.10, 0.25, 0.50, 0.75, 0.90\}$ ranges from low-inequality ($\tau=0.10$) to high-inequality ($\tau=0.90$) counties. Model~1 includes $PP\_TOTALEXP_{it}$ for total per-pupil expenditure. Model~2 uses $PP\_COMP_{k,it}$ representing per-pupil expenditure components: instruction ($PP\_TCURINST$), support services ($PP\_TCURSSVC$), other current operations ($PP\_TCURONON$), capital outlay ($PP\_TCAPOUT$), and interest payments ($PP\_TINTRST$). Model~3 includes $PS\_SHARE_{m,it}$ for spending shares: current expenditure share ($PS\_TCURELSC$), capital share ($PS\_TCAPOUT$), and interest share ($PS\_TINTRST$). Model~4 employs $PS\_DETAIL_{n,it}$ for detailed spending shares: instruction share ($PS\_TCURINST$), support services share ($PS\_TCURSSVC$), other current share ($PS\_TCURONON$), capital share ($PS\_TCAPOUT$), and interest share ($PS\_TINTRST$).

The control vector $\mathbf{X}_{it}$ includes economic indicators: poverty rate, unemployment rate, median household income, and annual percentage change in GDP; demographic variables are race and educational attainments. The term $\epsilon_{it}(\tau)$ represents the quantile-specific error.

\subsection{Quantile Regression Specification}
The quantile regression is employed to examine heterogeneous effects across different counties. This method estimates conditional quantile functions, allowing examination of how determinants of inequality vary across the entire distribution of the Gini coefficient rather than just at the conditional mean. Since the relationship between education spending and inequality may differ substantially between low-inequality and high-inequality counties, the quantile regression becomes the most appropriate estimation method to use as it captures the heterogeneous effect at every quantile.

Standard errors are calculated using the kernel method with Hall-Sheather bandwidth selection, which provides robust inference in the presence of heteroskedasticity.

\subsection{Methodological Considerations}
 First, the relationship between spending and inequality raises potential endogeneity concerns, as areas with different economic characteristics may have varying capacities for educational investment. While quantile regression does not resolve this identification challenge, it provides insights into how these relationships vary across different county contexts, offering a more nuanced understanding than mean-based approaches.

Second, to address measurement validity, all expenditure variables in Models 1 and 2 are expressed as per-pupil amounts, while Models 3 and 4 examine spending shares to understand budget composition effects. This dual approach examines both the level and allocation of educational resources, providing complementary perspectives on how education financing relates to inequality.

Third, the 12-year panel data (2010--2022) provides substantial temporal variation for examining medium-term relationships, though results should be interpreted with appropriate caution regarding long-run equilibrium effects.

Finally, Household median income, number of people Unemployed, and the number of people living below the poverty line are appropriately transformed using the inverse hyperbolic sine to address skewness and facilitate interpretation. Demographic variables (race, education levels) are expressed as percentages, and annual change in GDP is expressed in percentages.  All models include a comprehensive set of controls identified as theoretically important in the literature review.

\subsection{Why Education Impact Income Distribution}
\label{sec:theory}
Public education can affect income inequality through two main channels: by raising skills across the population and by changing the returns to different types of human capital. Following \citep{schultz1963}, education increases productivity and earnings. When financed publicly, it can weaken the link between parental resources and children's outcomes, as modeled by \citet{glomm1992, glomm2003}:

\begin{equation}
h_{t+1} = A(e_t + E_t)^\alpha h_t^{1-\alpha}
\end{equation}

where $h_{t+1}$ is child human capital, $E_t$ is public spending, and $h_t$ is parental human capital. More progressive $E_t$ reduces intergenerational inequality.

However, the net effect is ambiguous. \citet{artige2023} show that spending can simultaneously help lower-skilled workers (bottom-up) and disproportionately benefit managers (top-down). This creates heterogeneous effects that depend on initial inequality levels, motivating our quantile regression specification across $\tau \in \{0.10, 0.25, 0.50, 0.75, 0.90\}$.

Additionally, how money is spent may matter as much as how much is spent. Budget composition whether dollars go to instruction, support services, or capital outlay likely mediates the spending-inequality relationship. This justifies our four-model approach that examines total expenditures, disaggregated components, and budget shares.

Thus, we test whether education spending affects inequality differently in low- versus high-inequality counties, and whether these effects vary by spending category. This extends prior work by \citet{sylwester2002, goodspeed2000} from asking if spending matters to understanding where and how it matters most.

\subsection{Data Sources and Variable Construction}
The study employs a  multi-source, longitudinal dataset constructed by merging several publicly available datasets at the county level from 2010--2022. Table \ref{tab:data_sources} provides the sources and justification for each variable inclusion.
\begin{table}[ht]
\centering
\caption{Variable Definitions and Data Sources}
\label{tab:data_sources}
\small
\begin{tabular}{p{3cm}p{3.5cm}p{5.2cm}}
\hline
\textbf{Variable} & \textbf{Source} & \textbf{Code/Justification} \\
\hline
Gini Index & Census ACS 5-Year & B19083\_001: Standard inequality measure (0=equality, 1=inequality) \\

Expenditures & Census F-33 Survey & Comprehensive school finance data \\
Poverty Rate & Census ACS 5-Year & B17001\_002: Controls for area deprivation \\
Median Income & Census ACS 5-Year & B19013\_001: Captures local economic environment \\
Unemployment & BLS LAUS & Labor market health indicator \\
$\Delta$ Annual GDP (\%) & BEA & Controls for economic cycles \\
Race/Ethnicity & Census ACS 5-Year & B02001\_001: Controls for structural inequalities \\
Population & Census ACS 5-Year & B02001\_001: For per-capita calculations \\
Education Levels & Census ACS 5-Year & B15002: Adult educational attainment; determines local skill base \\
\hline
\end{tabular}
\end{table}


\begin{table}[ht]
\centering
\small
\caption{Descriptive Statistics}
\label{tab:descriptives}
\begin{tabular}{lrrrrrr}
\toprule
Variable & N & Mean & SD & Min & Median & Max \\
\midrule
\textbf{Income Inequality} \\
Gini Index & 184,686 & 0.442 & 0.034 & 0.340 & 0.440 & 0.652 \\
\hline
\textbf{Per-Pupil Spending (Thousands \$)} \\
Total Expenditure & 184,686 & 11.053 & 3.689 & 6.567 & 10.449 & 60.981 \\
Current Expenditure & 184,686 & 10.046 & 3.099 & 6.320 & 9.678 & 57.425 \\
\quad Instruction & 184,686 & 5.622 & 2.329 & 3.762 & 5.344 & 27.923 \\
\quad Support Services & 184,686 & 3.622 & 1.930 & 1.818 & 3.406 & 27.710 \\
\quad Other Current & 184,686 & 0.802 & 0.447 & 0.082 & 0.754 & 3.363 \\
Capital Outlay & 184,686 & 0.834 & 0.983 & 0.001 & 0.481 & 20.700 \\
Interest & 184,686 & 0.155 & 0.189 & 0.000 & 0.104 & 2.691 \\
\hline
\textbf{Spending Shares} \\
Current Share & 184,686 & 0.917 & 0.067 & 0.381 & 0.939 & 0.997 \\
\quad Instruction Share  & 184,686 & 0.515 & 0.063 & 0.194 & 0.525 & 0.672 \\
\quad Support Services  & 184,686 & 0.328 & 0.055 & 0.126 & 0.327 & 0.521 \\
\quad Other Current & 184,686 & 0.074 & 0.028 & 0.005 & 0.072 & 0.221 \\
Capital Outlay  & 184,686 & 0.068 & 0.049 & 0.00009 & 0.046 & 0.619 \\
Interest Share  & 184,686 & 0.014 & 0.014 & 0.00001 & 0.010 & 0.216 \\
\hline
\textbf{Demographic Characteristics (\%)} \\
White & 184,686 & 83.88 & 19.08 & 10.86 & 90.00 & 99.48 \\
Black & 184,686 & 10.16 & 15.54 & 0.006 & 3.129 & 87.79 \\
Asian & 184,686 & 1.035 & 1.971 & 0.004 & 0.587 & 52.228 \\
\hline
\textbf{Education Levels (\%)} \\
High School or Less & 184,686 & 34.082 & 9.667 & 6.457 & 34.680 & 56.778 \\
Some College & 184,686 & 20.382 & 3.683 & 7.345 & 20.260 & 37.796 \\
Bachelor's or Higher & 184,686 & 14.034 & 6.544 & 3.807 & 12.791 & 56.346 \\
\hline
\textbf{Economic Indicators} \\
Poverty Rate & 184,686 & 9.270 & 1.968 & 4.543 & 9.185 & 13.760 \\
Unemployment Rate & 184,686 & 7.438 & 1.781 & 2.893 & 7.302 & 12.743 \\
Household Median Income (log) & 184,686 & 11.46 & 0.252 & 10.57 & 11.45 & 12.54 \\
Annual GDP Growth (\%) & 184,686 & 1.441 & 5.715 & -44.100 & 1.400 & 113.500 \\
\hline
\textbf{School Characteristics} \\
Enrollment & 184,686 & 12,801 & 33,899 & 200 & 3,307 & 330,225 \\
\bottomrule
\end{tabular}
\smallskip
\\
\footnotesize \textit{Note:} Statistics based on county-year observations from 2010-2022 (N = 184,686). Per-pupil spending in thousands of dollars. Spending shares represent proportions of total budget allocated to each category (sum to 100\% within Current Expenditure group). Demographic and education variables expressed as percentages. Log-transformed variables indicated.
\end{table}

\section{Results} 
The descriptive statistics (Table \ref{tab:descriptives}) shows variation in both educational expenditures and inequality across U.S. counties from 2010--2022. The Gini index averages 0.442 with a standard deviation of 0.034. This indicates a moderate income inequality with meaningful cross-county variation. The minimum value of 0.340 represents relatively equal counties, while the maximum of 0.652 corresponds to highly unequal areas.

Per-pupil education spending averages \$11,053 (SD = \$3,689), ranging from \$6,567 to \$60,981. This highlights the disparities in educational resources across counties. Instruction comprises the largest component at \$5,622 per pupil on average, representing 51.5\% of current expenditures, while support services account for \$3,622 (32.8\%). Capital outlays and interest payments are relatively small shares of total expenditure at 6.8\% and 1.4\% respectively.

In the demographic composition, the white population constituted the majority of the racial distribution with and average of 83.9\% ranging from 10.9\% to 99.5\%, blacks constituted and average of 10.16\% and asians 1.035\%. The Educational attainment averaged 34\% for those who have a high school diploma or less, 20.4\% for those with some college and 14\% for bachelor's degree rates.

The median household income (inverse hyperbolic transform) shows moderate variation, while GDP growth rates exhibit substantial volatility with extreme values (-44.1\% to 113.5\%) likely reflecting localized economic shocks and recoveries during the sample period.

The results in Table \ref{tab:qr1} show that per-pupil education spending has a small but consistent positive relationship with income inequality across all counties. In every quantile, higher spending is linked with a slightly higher Gini Index. The effect grows stronger as we move from low-inequality counties (0.10 quantile) to high-inequality counties (0.90 quantile. In simple terms, this means that if a county increases its per-pupil spending, we would expect its Gini Index to rise a little. The increase is not large, but it is noticeable in already unequal counties. This reflects the fact that counties able to spend more on education also tend to be places with deeper income gaps.

The control variables also show patterns with clear implications. Poverty has a strong positive association with inequality at the lower quantiles, meaning that in more equal counties, a rise in poverty quickly widens income gaps. However, at the highest inequality levels, the effect becomes negative, suggesting that in very unequal places, increases in poverty slightly compress the distribution because many residents are already concentrated at the lower end.

Unemployment shows a small negative effect across all quantiles. This means that if unemployment increases, inequality tends to shrink slightly, likely because income losses are spread across many groups and reduce differences between them. Median household income has a strong negative effect everywhere. A rise in median income leads to a meaningful reduction in inequality, reflecting a stronger middle class that pulls the distribution together. This is one of the strongest relationships in the table.

The demographic and education variables also have predictable implications. A higher share of White residents is linked with lower inequality across all quantiles. For Black and Asian populations, the effects vary depending on the level of inequality, which shows that the relationship between race and income distribution is more complex and context-dependent. In terms of education levels, counties with more residents who have only a high school education tend to become more unequal as their share grows. At the same time, counties with more residents holding a bachelor’s degree or higher also experience higher inequality, which fits common labor-market patterns: places with many highly skilled workers often have wider income gaps.

The results in Table \ref{tab:qr2} help explain why total per-pupil spending in Table \ref{tab:qr1} shows a small but consistent positive relationship with income inequality. When we break spending into its components, we see that not all types of education spending work the same way, and some even move inequality in opposite directions. Instructional spending which is the largest category has the strongest positive relationship with inequality across all quantiles. The size of the coefficient grows sharply from low-inequality counties to high-inequality counties, rising from 0.00274 to 0.01052. This suggests that places able to spend more on classroom instruction tend to be richer counties where income gaps are already wider. In practical terms, an increase in instructional spending pushes inequality up slightly, but the effect is stronger in counties that are already unequal.

Support services show a smaller positive effect, meaning that higher spending on counseling, administration, transportation, and similar services is also associated with higher inequality, though the size of the effect is much weaker than for instructional spending. On the other hand, “other current spending,” which includes food services, Enterprise operations, and other elementary-secondary programs, has a negative effect at all quantiles. This means that when counties put relatively more money into these areas, inequality tends to fall a little. Capital outlay is mixed: it is positive at the lower quantiles but negative at the middle and upper quantiles, suggesting that construction and facilities spending may widen gaps in low-inequality counties but slightly compress the distribution in richer, high-spending areas. Interest payments consistently reduce inequality, with relatively large negative coefficients. This implies that counties with higher debt service burdens may be those with lower capacity to spend, which aligns more closely with lower inequality patterns.

The control variables in Table \ref{tab:qr2} behave almost exactly as they did in Table \ref{tab:qr1}, reinforcing the earlier interpretation. Poverty increases inequality at the lower end of the distribution but decreases it slightly in counties already at the top. Unemployment still slightly reduces inequality, and median household income remains one of the strongest equalizing forces in the model. The demographic and education variables also maintain the same general patterns, showing consistent relationships across the two models.

The results in Table \ref{tab:qr3} show that the way counties allocate their education budgets is strongly linked with income inequality, and these effects are much larger than those found for total per-pupil spending. Because the variables here are shares, the coefficients tell us what happens when counties shift one percentage point of their education spending from one category to another. Using the average per-pupil spending of about \$11,000, a one percentage-point shift represents roughly \$110 per student being reallocated.

Instructional salaries, school operations, and support services have a clear negative effect on inequality across all quantiles. At the median, the coefficient is about –0.123, meaning that if a county reallocates 1\% of its education budget (about \$110 per student) toward current operations and away from other categories, its Gini Index falls by roughly 0.0012 points. While this is a small change, it is meaningful when applied across an entire county population. The effect is even stronger in high-inequality counties, where the coefficient reaches –0.224, suggesting that a similar 1\% shift lowers the Gini by more than 0.0022 points.

Capital outlay and interest shares show similar patterns. A one percentage-point increase in capital outlay reduces the Gini Index by about 0.0013 at the middle of the distribution, and even more in high-inequality counties. The strongest effect comes from interest payments. Raising the interest share by just one percentage point reduces inequality by 0.0032 points in the highest quantile. 

Table \ref{tab:qr4} breaks down spending shares further, separating current spending into instruction, support services, and other current spending. The results show that reallocating education dollars toward any of these categories is associated with lower income inequality, and the effects are larger than those observed when using total per-pupil spending.
An increase in the instruction share by one percentage point reduces the Gini Index by about 0.0011 at the median, with a larger effect in high-inequality counties (–0.0019 at the 90th percentile). Support services behave similarly, and “other current spending” has an even stronger negative effect, especially in high-inequality counties, suggesting that funding smaller operational items like school meals or student services helps reduce gaps. Capital and interest shares continue to show strong inequality-reducing effects; a one-percentage point shift toward these areas decreases the Gini by up to 0.0035 at the top quantile.

The economic and demographic controls remain consistent with earlier models. Poverty increases inequality in low-inequality counties but has little or slightly negative effects in the most unequal areas. Higher median income consistently reduces inequality, while unemployment slightly compresses the distribution. Counties with more White residents tend to have lower inequality, while higher shares of college-educated residents are associated with higher inequality which may be linked to skill-based earnings differences.

Overall, the four models together show that both the level and composition of education spending matter for income inequality, but in different ways. Model \ref{eq:m3} and Model \ref{eq:m4} demonstrate that how counties allocate their budgets toward instruction, support services, and other current spending matter. It has a larger and more consistent effect on reducing inequality than simply increasing total per-pupil spending, as seen in Model \ref{eq:m1} and Model \ref{eq:m2}. Small reallocations of about \$100–\$150 per student from capital or debt payments to operational and instructional spending can meaningfully reduce inequality, particularly in high-inequality counties. Across all models, key economic and demographic factors: poverty, median income, and education levels have much larger effects on the Gini Index than spending alone. These findings highlight that targeted and strategic use of education funds is crucial for promoting equity, and that total spending levels without attention to allocation may hinder important opportunities to reduce disparities.

\section{Conclusions}
This study examined the relationship between public education spending and income inequality across U.S. counties, using four quantile regression models to account for variation across the distribution of inequality. The results show that total per-pupil spending has a small but consistent effect on the Gini Index, confirming that it is a justified predictor. However, the composition of spending matters far more than the total amount. Instructional and support service expenditures are the most effective at reducing inequality, while capital outlays and interest payments have mixed or smaller effects. Even relatively small reallocations of \$100–\$150 per student toward operational and instructional spending can meaningfully reduce inequality, especially in counties with the highest income gaps.

Economic and demographic factors such as poverty rates, median household income, and educational attainment have substantially larger impacts on inequality than changes in spending alone, highlighting the importance of broader social and economic contexts. Looking at how education funds are divided and spent, this study shows that carefully targeting specific types of spending can improve fairness and opportunities more effectively than just increasing total budgets.

Overall, the results show that both funding decisions and spending choices matter. By prioritizing instructional and support services and managing capital and debt wisely, education budgets can help reduce income inequality while supporting broader social and economic goals.

\section*{Appendix}
\begin{table}[htbp]
\centering
\small
\caption{Quantile Regression Results for Gini Index Determinants}
\label{tab:qr1}
\begin{tabular}{lccccc}
\toprule
 & \multicolumn{5}{c}{\textbf{Dependent Variable: Gini Index}} \\
\cmidrule(l){2-6}
\textbf{Variable} & \textbf{0.10} & \textbf{0.25} & \textbf{0.50} & \textbf{0.75} & \textbf{0.90} \\
\midrule
\textbf{Education Spending} \\
Per-Pupil Spending & 0.00085*** & 0.00113*** & 0.00152*** & 0.00199*** & 0.00241*** \\
\hline
\textbf{Economic Indicators} \\
Poverty Rate & 0.00894*** & 0.00672*** & 0.00342*** & 0.00073*** & -0.00281*** \\
Unemployment Rate & -0.00165*** & -0.00179*** & -0.00140*** & -0.00157*** & -0.00154*** \\
Household Median Income & -0.07472*** & -0.08313*** & -0.09182*** & -0.09858*** & -0.09705*** \\
Annual GDP Change (\%) & -0.00003** & -0.00004*** & -0.00002* & -0.00003* & 0.00000 \\
\hline
\textbf{Demographics} \\
White & -0.00062*** & -0.00062*** & -0.00059*** & -0.00068*** & -0.00081*** \\
Black & -0.00010*** & -0.00005** & 0.00004** & 0.00002 & 0.00003 \\
Asian & -0.00111*** & -0.00057*** & 0.00084*** & 0.00159*** & 0.00055** \\
\hline
\textbf{Education Attainment} \\
High School & 0.00102*** & 0.00096*** & 0.00087*** & 0.00080*** & 0.00056*** \\
Some College & 0.00043*** & 0.00034*** & 0.00033*** & -0.00006 & -0.00124*** \\
Bachelors Or Higher & 0.00318*** & 0.00334*** & 0.00362*** & 0.00390*** & 0.00392*** \\
\hline
\textbf{Constant} & 1.15435*** & 1.28267*** & 1.41570*** & 1.54385*** & 1.61553*** \\
\midrule
Observations & \multicolumn{5}{c}{184,686} \\
\bottomrule
\end{tabular}
\medskip
\footnotesize 
\\
\textit{Notes: Significance levels; *p$<0.05$, $**p<0.01$, $***p<0.001$. Quantiles represent points in the conditional Gini index distribution: $\tau=0.10$ (low inequality), $\tau=0.25$ (first quartile), $\tau=0.50$ (median), $\tau=0.75$ (third quartile), $\tau=0.90$ (high inequality).}
\end{table}

\begin{table}[htbp]
\centering
\small
\setlength{\tabcolsep}{4pt}
\caption{Quantile Regression: Disaggregated Education Spending Components- Model 2}
\label{tab:qr2}
\begin{tabular}{lccccc}
\toprule
 & \multicolumn{5}{c}{\textbf{Dependent Variable: Gini Index}} \\
\cmidrule(l){2-6}
\textbf{Variable} & \textbf{0.10} & \textbf{0.25} & \textbf{0.50} & \textbf{0.75} & \textbf{0.90} \\
\midrule
\textbf{Education Spending (Per-Pupil)} \\
Instruction & 0.00274*** & 0.00336*** & 0.00454*** & 0.00689*** & 0.01052*** \\
Support Services & 0.00075*** & 0.00176*** & 0.00168*** & 0.00128*** & 0.00089*** \\
Other Current & -0.00072** & -0.00326*** & -0.00224*** & -0.00296*** & -0.00840*** \\
Capital Outlay & 0.00018** & 0.00034*** & -0.00016** & -0.00028*** & -0.00015 \\
Interest payment & -0.01279*** & -0.01452*** & -0.00845*** & -0.00598*** & -0.00728*** \\
\hline
\textbf{Economic Indicators} \\
Poverty Rate & 0.00911*** & 0.00682*** & 0.00355*** & 0.00093*** & -0.00231*** \\
Unemployment Rate & -0.00154*** & -0.00159*** & -0.00125*** & -0.00136*** & -0.00127*** \\
 Household Median Income & -0.07793*** & -0.08606*** & -0.09524*** & -0.10321*** & -0.10637*** \\
Annual GDP Change (\%) & -0.00003* & -0.00003*** & -0.00002* & -0.00003** & -0.00001 \\
\hline
\textbf{Demographics} \\
White & -0.00064*** & -0.00058*** & -0.00054*** & -0.00063*** & -0.00066*** \\
Black & -0.00013*** & -0.00004** & 0.00007*** & 0.00004*** & 0.00014*** \\
Asian & -0.00124*** & -0.00025** & 0.00104*** & 0.00172*** & 0.00128*** \\
\hline
\textbf{Education Attainment} \\
High School & 0.00095*** & 0.00078*** & 0.00072*** & 0.00068*** & 0.00055*** \\
Some College & 0.00034*** & 0.00017*** & 0.00017*** & -0.00012*** & -0.00111*** \\
Bachelors or Higher & 0.00314*** & 0.00313*** & 0.00345*** & 0.00380*** & 0.00403*** \\
\hline
\textbf{Constant} & 1.18928*** & 1.31461*** & 1.44697*** & 1.57775*** & 1.66945*** \\
\midrule
Observations & \multicolumn{5}{c}{184,686} \\
\bottomrule
\end{tabular}
\medskip
\footnotesize 
\\
\textit{Notes: Significance levels; *p$<0.05$, $**p<0.01$, $***p<0.001$. Quantiles represent points in the conditional Gini index distribution: $\tau=0.10$ (low inequality), $\tau=0.25$ (first quartile), $\tau=0.50$ (median), $\tau=0.75$ (third quartile), $\tau=0.90$ (high inequality).}
\end{table}

\begin{table}[htbp]
\centering
\small
\setlength{\tabcolsep}{4pt}
\caption{Quantile Regression: Education Spending Shares and Inequality -Model 3}
\label{tab:qr3}
\begin{tabular}{lccccc}
\toprule
 & \multicolumn{5}{c}{\textbf{Dependent Variable: Gini Index}} \\
\cmidrule(l){2-6}
\textbf{Variable} & \textbf{0.10} & \textbf{0.25} & \textbf{0.50} & \textbf{0.75} & \textbf{0.90} \\
\midrule
\textbf{ Education Spending Shares} \\
Current Expenditure & -0.11929*** & -0.16640*** & -0.12335*** & -0.18121*** & -0.22440*** \\
Capital Outlay & -0.12184*** & -0.16414*** & -0.12601*** & -0.18708*** & -0.22399*** \\
Interest Share & -0.23119*** & -0.28278*** & -0.18973*** & -0.26884*** & -0.32211*** \\
\hline
\textbf{Economic Indicators} \\
Poverty Rate & 0.00884*** & 0.00679*** & 0.00328*** & 0.00065*** & -0.00295*** \\
Unemployment Rate & -0.00175*** & -0.00185*** & -0.00157*** & -0.00174*** & -0.00161*** \\
Household Median Income & -0.07473*** & -0.08122*** & -0.08776*** & -0.09312*** & -0.09220*** \\
Annual GDP Change (\%)& -0.00003* & -0.00004*** & -0.00002* & -0.00002 & -0.00001 \\
\hline
\textbf{Demographics} \\
White & -0.00064*** & -0.00066*** & -0.00064*** & -0.00078*** & -0.00092*** \\
Black & -0.00011*** & -0.00010*** & 0.00001 & -0.00007*** & -0.00003 \\
Asian & -0.00122*** & -0.00073*** & 0.00058*** & 0.00144*** & 0.00026 \\
\hline
\textbf{Education Attainment} \\
High School & 0.00103*** & 0.00100*** & 0.00097*** & 0.00091*** & 0.00069*** \\
Some College & 0.00047*** & 0.00037*** & 0.00040*** & 0.00004 & -0.00104*** \\
Bachelors or Higher & 0.00322*** & 0.00339*** & 0.00370*** & 0.00398*** & 0.00403*** \\
\hline
\textbf{Constant} & 1.28662*** & 1.44227*** & 1.51149*** & 1.69078*** & 1.81448*** \\
\midrule
Observations & \multicolumn{5}{c}{184,686} \\
\bottomrule
\end{tabular}
\medskip
\footnotesize 
\\
\textit{Notes: Significance levels; *p$<0.05$, $**p<0.01$, $***p<0.001$. Quantiles represent points in the conditional Gini index distribution: $\tau=0.10$ (low inequality), $\tau=0.25$ (first quartile), $\tau=0.50$ (median), $\tau=0.75$ (third quartile), $\tau=0.90$ (high inequality).}
\end{table}

\begin{table}[htbp]
\centering
\small
\setlength{\tabcolsep}{4pt}
\caption{Quantile Regression: Disaggregated Spending Shares - model 4}
\label{tab:qr4}
\begin{tabular}{lccccc}
\toprule
 & \multicolumn{5}{c}{\textbf{Dependent Variable: Gini Index}} \\
\cmidrule(l){2-6}
\textbf{Variable} & \textbf{0.10} & \textbf{0.25} & \textbf{0.50} & \textbf{0.75} & \textbf{0.90} \\
\midrule
\textbf{Current Spending Shares} \\
Instruction & -0.11892*** & -0.16999*** & -0.11251*** & -0.16860*** & -0.18606*** \\
Support Services & -0.12363*** & -0.17034*** & -0.10354*** & -0.16517*** & -0.19399*** \\
Other Current Spending & -0.16201*** & -0.24748*** & -0.19884*** & -0.26935*** & -0.33304*** \\
\hline
\textbf{Capital \& Interest Shares} \\
Capital & -0.12691*** & -0.17309*** & -0.11746*** & -0.18029*** & -0.20157*** \\
Interest & -0.25780*** & -0.33311*** & -0.21900*** & -0.30499*** & -0.34720*** \\
\hline
\textbf{Economic Indicators} \\
Poverty Rate & 0.00885*** & 0.00669*** & 0.00331*** & 0.00080*** & -0.00241*** \\
Unemployment Rate & -0.00170*** & -0.00177*** & -0.00144*** & -0.00169*** & -0.00156*** \\
Household Median Income & -0.07508*** & -0.08115*** & -0.08794*** & -0.09415*** & -0.09544*** \\
Annual GDP change (\%) & -0.00002* & -0.00004*** & -0.00002 & -0.00002 & 0.00000 \\
\hline
\textbf{Demographics} \\
White & -0.00065*** & -0.00065*** & -0.00063*** & -0.00077*** & -0.00088*** \\
Black & -0.00012*** & -0.00008*** & 0.00002 & -0.00007*** & -0.00002 \\
Asian & -0.00115*** & -0.00067*** & 0.00050*** & 0.00123*** & 0.00000 \\
\hline
\textbf{Education Attainment} \\
High School& 0.00104*** & 0.00092*** & 0.00094*** & 0.00087*** & 0.00064*** \\
Some College & 0.00044*** & 0.00022*** & 0.00033*** & -0.00003 & -0.00109*** \\
Bachelors or Higher & 0.00323*** & 0.00331*** & 0.00366*** & 0.00394*** & 0.00403*** \\
\hline
\textbf{Constant} & 1.29620*** & 1.45807*** & 1.50730*** & 1.69764*** & 1.82221*** \\
\midrule
Observations & \multicolumn{5}{c}{184,686} \\
\bottomrule
\end{tabular}
\medskip
\footnotesize
\\
\textit{Notes: Significance levels; *p$<0.05$, $**p<0.01$, $***p<0.001$. Quantiles represent points in the conditional Gini index distribution: $\tau=0.10$ (low inequality), $\tau=0.25$ (first quartile), $\tau=0.50$ (median), $\tau=0.75$ (third quartile), $\tau=0.90$ (high inequality).}
\end{table}

\begin{figure}[htbp]
    \centering
    \includegraphics[width=0.80\linewidth]{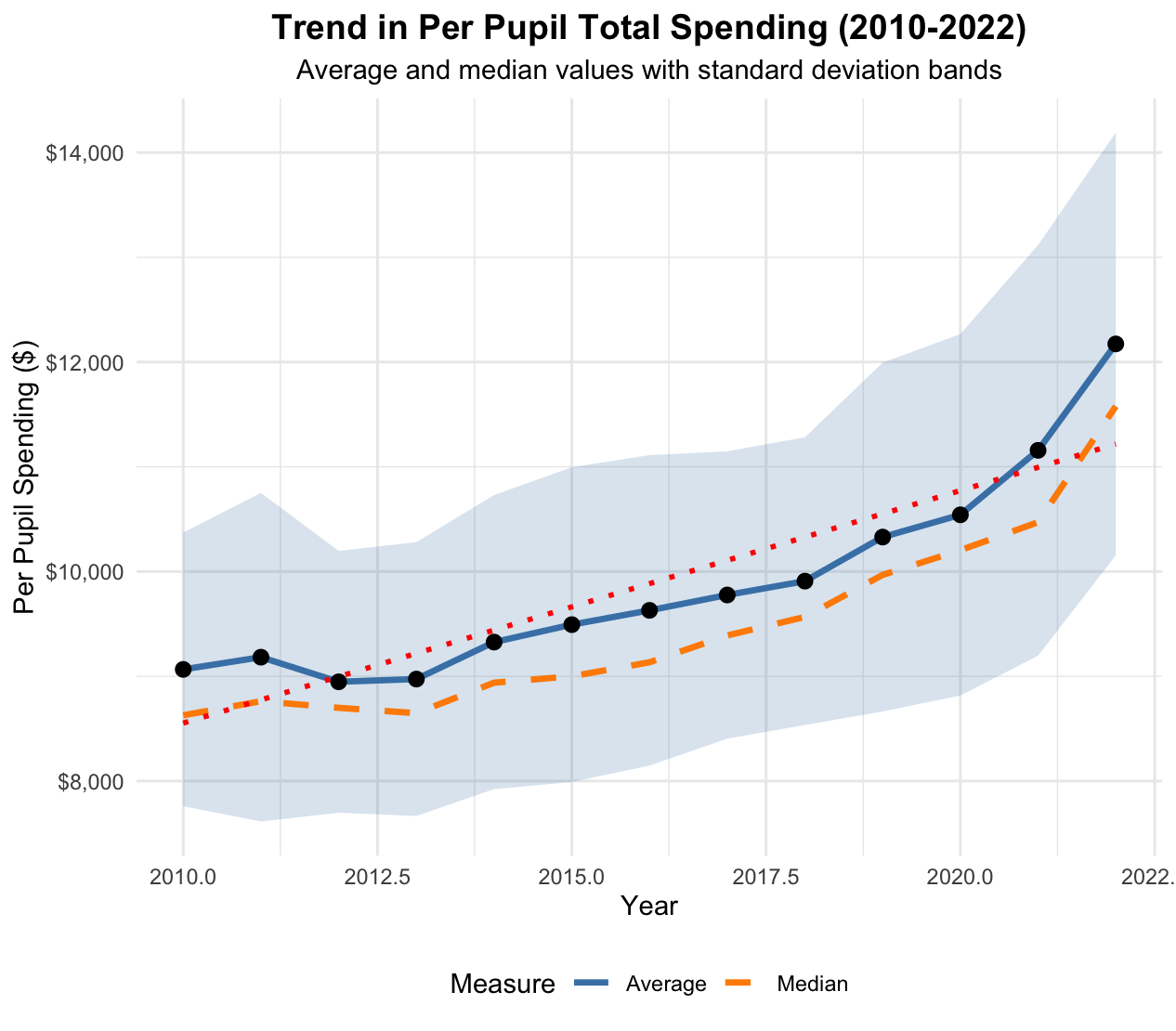}
    \caption{Trends in Per-Pupil Total Spending (2010-2022)}
    \label{fig:placeholder}
\end{figure}

\begin{figure}[htbp]
    \centering
    \includegraphics[width=0.80\linewidth]{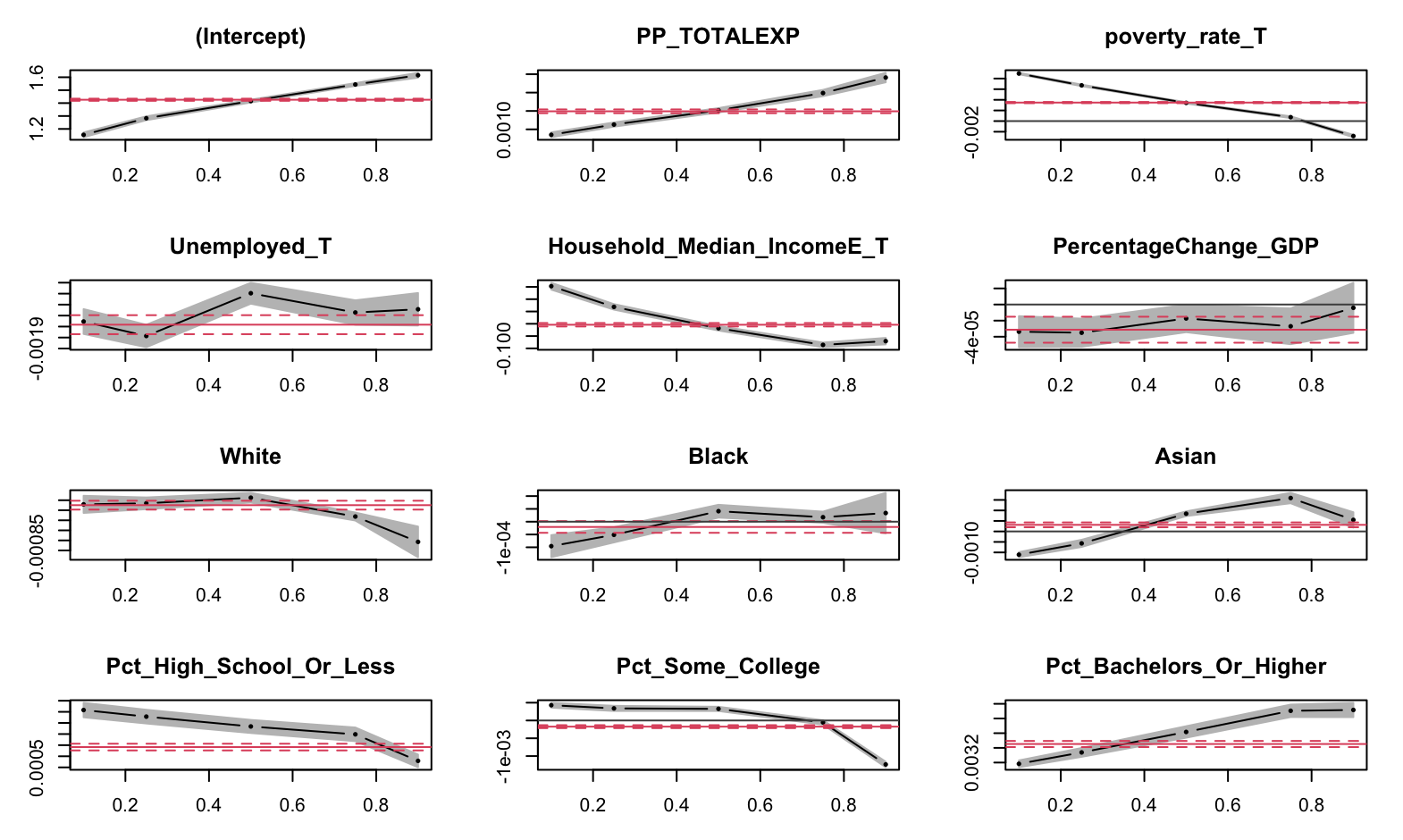}
    \caption{Quantile Regression plot for model 1}
    \label{fig:qr1}
\end{figure}

\begin{figure}[ht]
    \centering
    \includegraphics[width=0.80\linewidth]{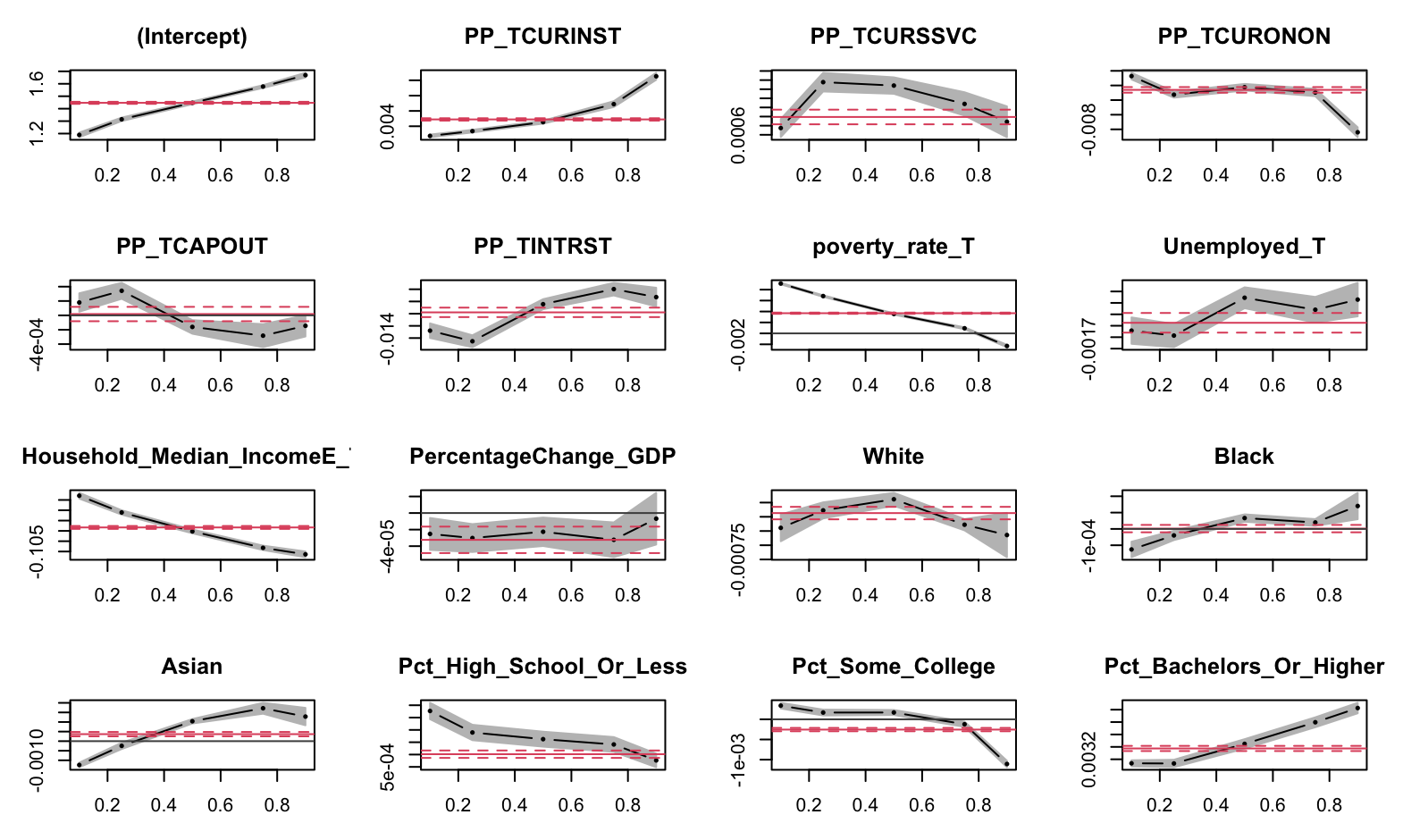}
    \caption{Quantile Regression plot for model 2}
    \label{fig:qr2}
\end{figure}

\begin{figure}[ht]
    \centering
    \includegraphics[width=0.80\linewidth]{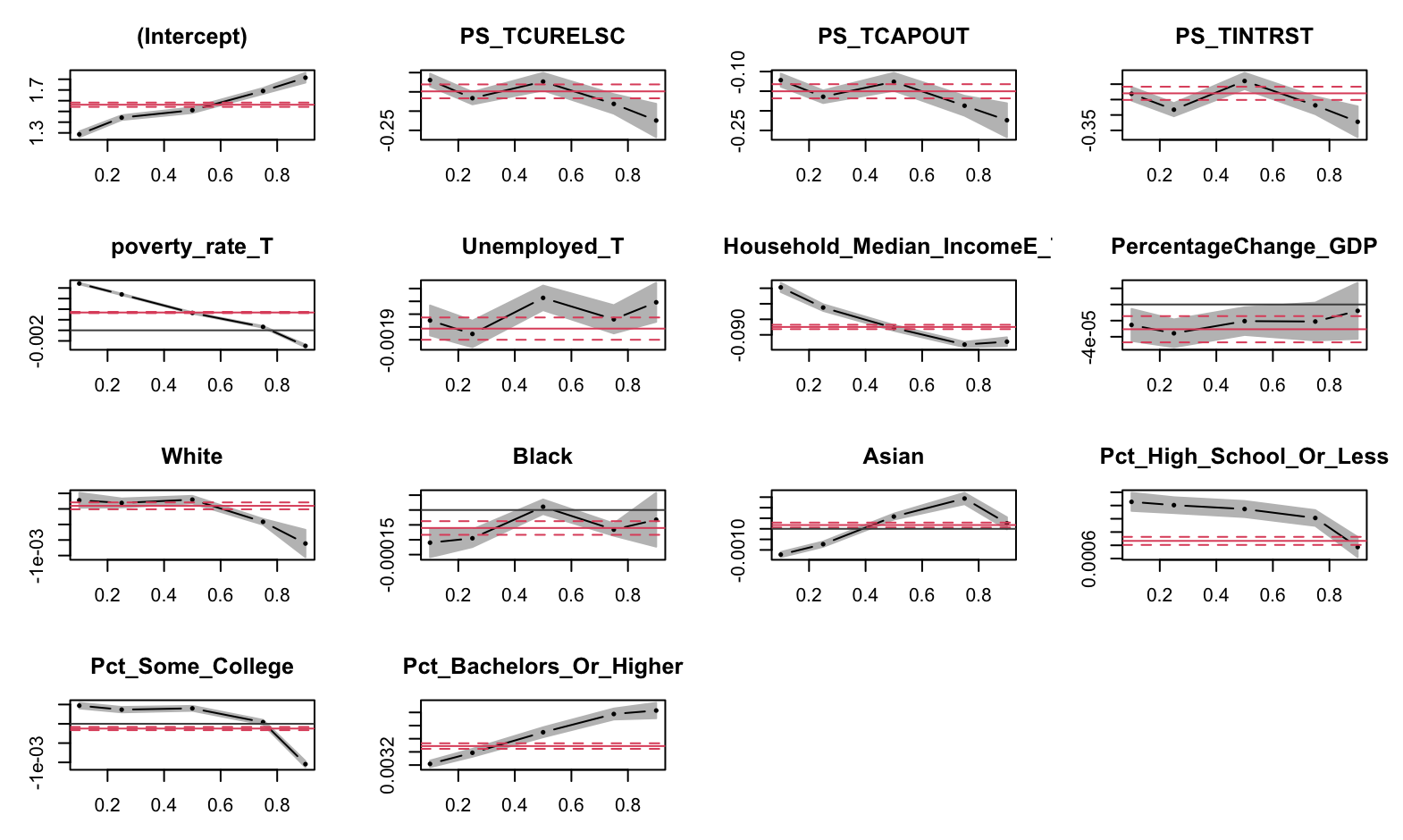}
    \caption{Quantile Regression plot for model 3}
    \label{fig:qr3}
\end{figure}

\begin{figure}[ht]
    \centering
    \includegraphics[width=0.80\linewidth]{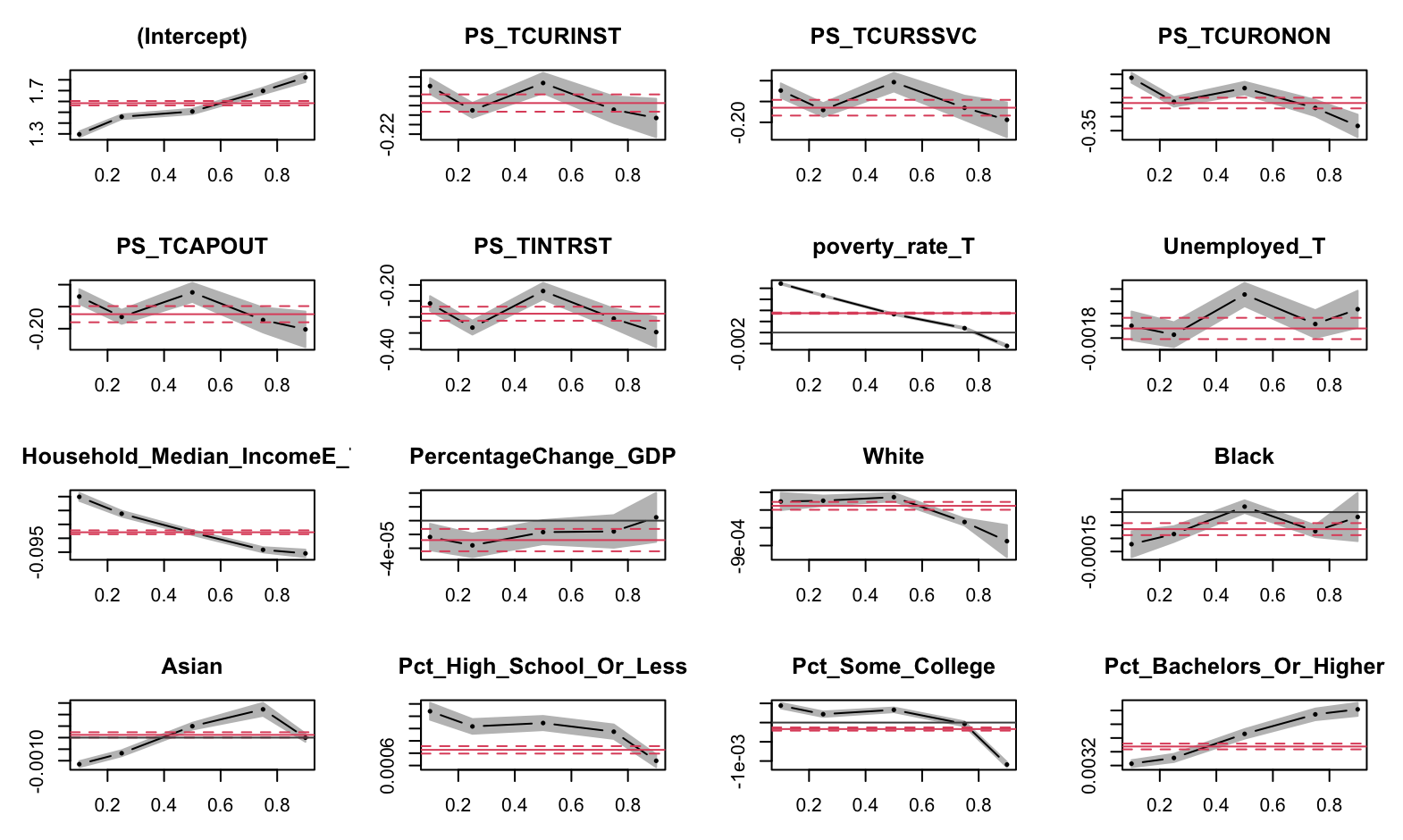}
    \caption{Quantile Regression plot for model 4}
    \label{fig:qr4}
\end{figure}
\clearpage
\bibliographystyle{apalike} 
\bibliography{my_bibliography_file} 

\end{document}